\documentstyle[multicol,aps,prl,twocolumn,psfig]{revtex}

\newif\iffigs
\figstrue   % uncomment if figures are present
\iffigs
\fi
\def\drawing #1 #2 #3 {
\begin{center}
\setlength{\unitlength}{1mm}
\begin{picture}(#1,#2)(0,0)
\put(0,0){\framebox(#1,#2){#3}}
\end{picture}
\end{center} }

\def\la{\left\langle}
\def\ra{\right\rangle}
\def\lb{\left\{}
\def\rb{\right\}}
\def\rf#1{(\ref{#1})}

\def\xsc#1{Section~\ref{#1}}

\def\xfg#1{Fig.~\ref{#1}}
\def\hmin{h_{\rm min}}
\def\hmax{h_{\rm max}}
\def\prob{{\rm Prob}\,}
\def\da{D^\alpha}
\def\hs{h_\star}
\def\ps{{p_\star}}

\def\mo{m^{(1)}}
\def\md{m^{(2)}}
\def\wu{w^{(1)}}
\def\wt{w^{(2)}}
\def\rset{{\rm I\kern -0.2em R}}
\def\un{\hbox{{1\kern -0.25em\raise 0.4ex\hbox{{\scriptsize $|$}}}}}

\def\cset{eq_burg_force\hbox{{C\kern -0.55em\raise 0.5ex\hbox{{\tiny $|$}}}}}
\def\nset{\hbox{{I\kern -0.18em N}}}

\def\eql{\,{\stackrel{\rm law}{=}}\,}

\begin{document}

%if twocolumns

\twocolumn[\hsize\textwidth\columnwidth\hsize\csname@twocolumnfalse\endcsname

\title{On multifractality and fractional derivatives} \author{U. Frisch$^{\rm
a}$ and T. Matsumoto$^{\rm a,b}$} \address{$^{\rm a}$
CNRS UMR 6529, Observatoire de la C\^ote d'Azur, BP 4229, 06304 Nice
Cedex 4, France \\$^{\rm b}$ Dep. Physics, Kyoto University,
Kitashirakawa Oiwakecho Sakyo-ku, Kyoto 606-8502, Japan} \draft
\date{\today} \maketitle

\centerline{to appear in {\it J. Stat. Phys.} {\bf 108}: 1181--1202 (2002)}

\begin{abstract}
It is shown phenomenologically that the fractional derivative $\xi=D^\alpha u$ of order $\alpha$
of a multifractal function has a power-law tail $\propto |\xi| ^{-p_\star}$ in
its cumulative probability, for a suitable range of $\alpha$'s. The exponent
is determined by the condition $\zeta_{p_\star} = \alpha p_\star$, where
$\zeta_p$ is the exponent of the structure function of order $p$. A detailed
study is made for the case of random multiplicative processes (Benzi {\it et
al.}  1993 Physica D {\bf 65}: 352) which are amenable to both theory and
numerical simulations. Large deviations theory provides a concrete criterion,
which involves the departure from straightness of the $\zeta_p$ graph, for the
presence of power-law tails when there is only a limited range over which the
data possess scaling properties (e.g. because of the presence of a viscous
cutoff).  The method is also applied to wind tunnel data and financial data.

\end{abstract}
\vspace*{5mm}
]
\noindent{\em To David and Yasha}

\section{Introduction}
\label{s:intro}

Multifractality for functions was introduced by Parisi and
Frisch \cite{pf85} to interpret experimental results of Anselmet 
{\it et al.} 
on fully developed turbulence \cite{agha84}. 
We show in this paper that multifractality is connected to the 
tail behavior of fractional derivatives (to be defined precisely later). 
First, in \xsc{s:pheno} we give a  simple
phenomenological argument predicting power-law tails in probability
distributions of fractional derivatives and suggesting practical constraints
on their observability. In \xsc{s:rmp} we consider the case of a class of
synthetic multifractal functions, the random multiplicative processes 
of Benzi {\it et al.} 
\cite{benzi93}, for which theoretical results are obtained by relating them to
the theory of random linear maps \cite{kesten73}. In particular, using large
deviations theory, we can predict the number of generations in a given random
multiplicative process needed to observe power-laws tails in probabilities of
fractional derivatives. Numerical experiments with such processes are presented
in \xsc{s:numericrmp}. In \xsc{s:data} we analyze high-Reynolds number wind
tunnel turbulence data and financial data; we show that the former, having
only weak deviations from  self-similar behavior \`a la Kolmogorov 1941, are 
unlikely to display power-law tails at accessible Reynolds numbers.
\xsc{s:conclusion} gives the conclusions. In this paper we consider
exclusively multifractal functions; the case of multifractal measures,
arising, e.g., from chaotic dynamical systems or the dissipation of
turbulent flows (see, e.g., Ref~\cite{mandelbrot74}), will be addressed 
elsewhere.  

\section{Phenomenology} 
\label{s:pheno}

A (homogeneous random) function $u(x)$ with $x\in\rset^d$ is called
multifractal if (for some real interval $0<\hmin\le h \le \hmax < 1$) there is
a continuum of sets ${\cal S}_{h}$ of fractal dimension $D(h)$ such that, when
$x$ belongs to ${\cal S}_{h}$, the function $u$ has a H\"older-like behavior
with exponent $0<h < 1$ \cite{pf85}.  In terms of increments this is expressed
as
\begin{equation}
|\delta u(x,r)| \propto |r|^{h}, \quad \!\!r \to 0, \quad 
\!\!x \in {\cal S}_{h},
\label{defholder}
\end{equation}
where $\delta u(x, r) = u(x + r) - u(x)$.
If we take $D(h)$ to be the covering dimension (see Ref.~\cite{frisch95}
Section~8.5.1 and references therein) and use instead the codimension
$F(h) \equiv d -D(h)$, where $d$ is the dimension of space,  and then
``thicken'' the set ${\cal S}_{h}$ by covering it with balls of small radius
$\rho$, into the set ${\cal S}^{\rho}_h$, we have 
\begin{equation}
  \prob \{ x \in {\cal S}^{\rho}_h \} \propto 
    \rho^{F(h)},\quad \rho\to 0.
    \label{codim}
\end{equation}

For the case of an ergodic random function, this probability can also be
interpreted as a fraction of space.  Note that by using codimensions the
extension from the one-dimensional to the multi-dimensional case is
straightforward (just replace scalars used throughout this paper by vectors).
For more background on multifractals see Ref.~\cite{frisch95} and references
therein.

We turn now to fractional derivatives.  The definition used here for
the fractional derivative $\da u(x)$ of order $0 \le \alpha < 1$ of a
function $u(x)$ is to multiply its Fourier transform $\hat u(k)$ by
$|k|^\alpha$.  This might be called, more correctly, the fractional
(negative) Laplacian of order $\alpha/2$.  In physical space our
fractional derivative may be written, in the one-dimensional case, as
\begin{equation}
\!\!\!\da u(x) \equiv -{\alpha\over \pi}\sin{\alpha\pi\over 2}
\Gamma(\alpha){\rm P.V.}\int {u(x+r)-u(x)\over |r|^{\alpha+1}}\, dr,
\label{defxfrac}
\end{equation}
where $\Gamma(\alpha)$ is the Gamma function and P.V. denotes a Cauchy
principal value.  The Fourier-space definition is very convenient for periodic
functions.  When working with non-periodic experimental data, windowing may be
necessary before the Fourier transformation is applied. Note that other
definitions of the fractional derivative, for example that based on the
ordinary derivative of the the Riemann--Liouville fractional integral
\cite{fracderhistory}, do not produce substantially different results, as far
as our work is concerned.

It is clear from \rf{defxfrac} that, at a point $x_0$ at which $u(x)$ has
H\"older exponent $h>\alpha$ the fractional derivative $\da u(x_0)$ is finite
(provided the function $u$ is, say, bounded). If $\alpha\ge h$ the fractional
derivative will generally be infinite. This is also the case with the
definition of fractional derivatives used in Ref.~\cite{stiassnie97}.  But
what takes place in the {\it neighborhood\/} of such a point? Let us first
consider the case of an isolated non-oscillatory singularity of exponent $h$
at $x$. That is, we have $|u(x+r) -u(x)| \propto |r|^h$ for small $r$.  By
substitution into \rf{defxfrac} it is easily checked that, with a suitable
constant $B>0$
\begin{equation}
|\da u(y)| \sim B |y-x|^{h-\alpha},\quad y\to x,
\label{hminusalph}
\end{equation}
a relation which could have been guessed by simple dimensional analysis.
Since $h-\alpha<0$, the fractional derivative  becomes very large near $x$.

What happens if instead of isolated singularities we have multifractal
non-oscillatory singularities?  By a standard argument of multifractal
analysis, in the phenomenological presentation of Ref.~\cite{pf85} (see also
Ref.~\cite{frisch95} Section~8.5), we still use \rf{hminusalph} for
determining the contribution of points $x \in {\cal S}_h$ to $\da u(y)$
calculated at points which are near ${\cal S}_h$ without belonging to this
set. It is now very simple to estimate the probability to have $|\da u|>\xi$
for large positive $\xi$. From \rf{hminusalph}, the contribution from a given
$h<\alpha$ to $|\da u(y)|$ will be greater than $\xi$ provided that
\begin{equation}
|y-x| <\left({\xi\over B}\right)^{-{1\over {\alpha -h}}}.
\label{howclose}
\end{equation}
By \rf{codim}, this has probability
\begin{equation}
\prob \lb |\da u|>\xi\rb \propto |y-x|^{F(h)} \propto \xi ^{-{F(h)\over 
{\alpha -h}}}.
\label{nearlythere}
\end{equation}
This is a power law in $\xi$. As we vary the H\"older exponent $h$, the exponent
of this power law changes. When we sum the contributions of the various
$h$'s (with suitable regular weights $d\mu (h)$ which we need not specify) 
we get an integral which, when evaluated by Laplace's method, gives
us leading-order power-law behavior for the tail probability:
\begin{eqnarray}
&&\prob \lb |\da u|>\xi\rb \propto \xi ^{-p_\star},
\label{powerlaw}\\[1ex]
&&p_\star(\alpha) = \inf_{h<\alpha} {F(h)\over {\alpha -h}}.
\label{pstar}
\end{eqnarray}
 
We shall now show that $p_\star$ can be easily found from the 
scaling exponents $\zeta_p$  of structure functions. We remind the reader
that, when $u(x)$ is a homogeneous multifractal function, the moments of its
increments, called structure functions, are defined as
\begin{equation}  
S_p(r) \equiv \la \left(|\delta u(r)|\right)^p\ra, \quad p\ge 0.
\label{defsp}
\end{equation}
They follow power laws at small $r$'s:
\begin{equation}
S_p(r) \propto |r|^{\zeta_p}, 
\label{zetap}
\end{equation}
where the up convex function $\zeta_p$ and the down convex 
function $F(h)$ are Legendre--Fenchel transforms of each other 
\cite{pf85}\cite{frisch95}
\begin{equation}
F(h) = \sup_p\left(\zeta_p -ph\right),\quad \zeta_p = \inf_h\left( ph +F(h)\right).
\label{LF}
\end{equation}
Using\rf{LF} in \rf{pstar} and interchanging the orders of the infimum and 
the supremum, we obtain
\begin{equation}
\ps =\sup_{p\ge0}\,\inf_{h<\alpha} \,{\zeta_p -ph \over \alpha -h}.
\label{LFforps}
\end{equation}
The infimum over $h<\alpha$ is $-\infty$ if $\zeta_p-p\alpha<0$, it equals $p$
if $\zeta_p-p\alpha = 0$ and is less than this value if $\zeta_p-p\alpha
>0$. Thus $\ps(\alpha)$ is the supremum of all $p$'s such that
$\zeta_p-p\alpha\ge 0$; in other words, it is the solution of
\begin{equation}
\zeta_\ps = \ps(\alpha) \alpha,
\label{zetappalpha}
\end{equation}
which is unique in view of the convexity of $\zeta_p$ and  vanishing for 
$p=0$ (a consequence of \rf{defsp}).
Thus, the exponent $\ps(\alpha)$ of the
tail probability of the fractional derivative of order $\alpha$ can be obtained
from the graph of $\zeta_p$ by the simple construction shown in \xfg{f:zetap}:
a line through the origin of slope $\alpha$ intersects the graph at $p=\ps(\alpha)$.
\begin{figure}
\iffigs 
\centerline{\psfig{file=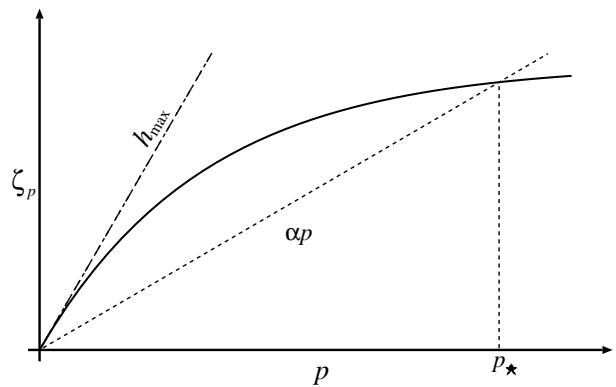,width=8cm}}
\else\drawing 65 10 {zetap}
\fi
\vspace{2mm}
\caption{Geometrical determination of the exponent $\ps(\alpha)$ of the power-law
tail from the graph of $\zeta_p$, the structure-function exponents. $h_{\rm
max}$ is the maximum H\"older exponent; $h_{\rm min}$, the minimum H\"older
exponent, is the (algebraically) smallest slope, not shown.}
\label{f:zetap}
\end{figure}

This construction has a number of interesting consequences which are now
presented. First, recall
that 
\begin{equation}
\hs = \left. {d\zeta_p \over dp} \right|_{p = \ps}.
\label{hszetap}
\end{equation}
(This follows from the observation that $F(h)$, if down convex, can be
recovered from $\zeta_p$ by another Legendre transformation.)
If there is a minimum H\"older exponent $\hmin$ and maximum exponent $\hmax$,
it is clear that $\alpha$ has to be between those two values for $\ps$ to
exist. If $\alpha \le \hmin$ we do not expect any power-law tail at all
(at least not by the mechanism considered here). If $\alpha > \hmax$ we expect
$\da u(x)$ to be infinite almost everywhere. $\da u(x)$ can then be made
finite if we assume that there is some ultraviolet cutoff, e.g. a viscous
cutoff. If so, the tail probability of $\da u(x)$ can be estimated by the
same kind of phenomenological arguments used for the tail probability of
velocity gradients in turbulence \cite{frisch95,fs91,bbpvv91}.

Next, we can use the construction to estimate how far the power-law tail (if
present at all) is expected to extend when multiscaling holds only over a
finite range of scales. By \rf{hminusalph} the fractional derivative stemming
from the H\"older exponent $\hs$ is given by
\begin{eqnarray}
&&|\da u(y)| \sim B |y-x|^{-C(\alpha)},\label{firstC}\\
&&C(\alpha)= \alpha -\hs = \left. {\zeta_{\ps} \over \ps}-{d\zeta_p\over dp}\right|_{_{p=\ps}}.
\label{defC}
\end{eqnarray}
$C(\alpha)$ will be called the {\it multifractality parameter}. In the next
sections we shall see that the presence of a power-law tail requires in practice
\begin{equation}
n C(\alpha) \ge 10,
\label{criterion}
\end{equation}
where $n$ is the number of octaves in the scaling range (e.g. in the inertial
range for turbulence). If the range of spatial scales is somewhat limited, it
is crucial that $C(\alpha)$ be as large as possible. In those instances where
the graph of $\zeta_p$ is close to a straight line through the origin, the two
terms on the r.h.s. of \rf{defC} will nearly cancel and $C(\alpha)$ will be very
small. Clearly, having a graph departing strongly from a straight line
will help in seeing power-law tails.

All the material presented in this phenomenological section is, by definition,
``soft''. Is it possible to give harder evidence? One way is to work with
Burgers turbulence in the limit of vanishing viscosity (Ref.~\cite{fb01} and
references therein). The singularities (mostly shocks) are then isolated and
it is easy to make our phenomenological arguments rigorous. Solutions to the
Burgers equation are however bifractal, not truly  multifractal.
In the next section we discuss a simple example of a truly multifractal
function.

\section{The case of the multifractal random multiplicative process}
\label{s:rmp}

In 1993 Benzi {\it et al.} \cite{benzi93} introduced an explicit method
for constructing multifractal random functions with arbitrary $\zeta_p$.\
Such functions are constructed by iterating suitable random maps and will
be called here ``random multiplicative processes'' (rmp). Our method of
construction differs somewhat from that of Ref.~\cite{benzi93} but is basically
equivalent. 

We first define the local rmp as a real random function on the real line. 
Let $\phi(x)$ be an indefinitely differentiable function with rapid decrease
at infinity and vanishing space integral. In practice we shall take
the Mexican hat function of width $\sigma$ used in Ref.~\cite{benzi93}
\begin{equation}
\phi(x)\equiv -{d^2\over dx^2} \exp\left(-{x^2\over 2\sigma ^2}\right).
\label{mexican}
\end{equation}
Let $m$ be a random variable, called the {\it random multiplier},  with symmetric distribution and finite moments of
all orders, characterized by its cumulative distribution
\begin{equation}
P(\eta) \equiv \prob \lb m>\eta\rb
\label{defPm}
\end{equation}
and its probability density
\begin{equation}
p(\eta)\equiv -{dP(\eta)\over d\eta} = p(-\eta).
\label{defpm}
\end{equation}
We shall also need independent identically distributed copies of $m$,
denoted  $\mo_n$ and  $\md_n$ ($n=0,1,\ldots$). (Henceforth the words
``identically distributed'' are understood.) We define now a sequence of 
random functions $w_n(x)$ recursively by
\begin{eqnarray}
&& w_0(x)=\phi(x), \label{defwz}\\
&& w_{n+1}(x)=\phi(x) +\mo_n\wu_n(2x)+\md_n\wt_n(2x-1), \label{defwn}\\
&&n=0,1,\ldots,\nonumber
\end{eqnarray}
where $\wu_n(x)$ and $\wt_n(x)$ are independent  copies
of $w_n(x)$.
The almost sure pointwise limit for $n\to\infty$ of $w_n(x)$,
if it exists, is denoted by $w(x)$ and satisfies obviously
\begin{equation}
w(x)\eql\phi(x) +\mo\wu(2x)+\md\wt(2x-1),
\label{weql}
\end{equation}
where $\eql$ designates equality in law.
It may be checked that our random map definition is equivalent to that
given in Ref.~\cite{benzi93}, which involves an infinite series.

Assuming the limit to exist, we now define the global rmp as
\begin{equation}
u(x) \equiv \sum_{i=-\infty}^{i=+\infty} w^{(i)}(x+x_i),
\label{defglobal}
\end{equation}
where the $w^{(i)}(x)$ are independent copies of $w(x)$ and the $x_i$ are
Poisson-distributed points on the real line with density $c$. 

The construction of the global rmp from the local one, which differs slightly
from that of Ref.~\cite{benzi93}, guarantees statistical homogeneity.
Denoting statistical means (expectation values) by angular brackets, we can
relate the characteristic functionals of $u(x)$ and $w(x)$.
Indeed, we have
\begin{eqnarray}
&&\la \exp\lb \int_\rset \imath\varphi(x)u(x)\,dx\rb\ra\nonumber\\
&&= \exp\lb c\int_\rset\left[\la\exp\left(\int_\rset\imath\varphi(x)w(x+y)\,dx\right)\ra-1\right]\,dy\rb, \nonumber \\
\label{baaaad}
\end{eqnarray}
where $\varphi(x)$ is a test function and $\int_\rset$ denotes integration
from $-\infty$ to $+\infty$. (To establish \rf{baaaad} it is convenient to
assume that there are $N$ points $x_i$ distributed uniformly in the interval
$[-L,+L]$ and then let $N\to\infty$, $L\to \infty$ and $N/(2L)\to c$.).
A special case of \rf{baaaad} relates just the characteristic functions (which
for $u$ does not depend on $x$):
\begin{equation}
\la e ^{\imath zu}\ra
= \exp\lb c\int_\rset\left(\la e ^{\imath zw(y)}\ra-1\right)\,dy\rb.
\label{good}
\end{equation}

The fractional derivative $\da u(x)$ is obtained by replacing,
in \rf{defglobal}, $w(x)$ by $\da w(x)$. Applying $\da$ to \rf{weql}, we find
that $\da w(x)$ satisfies
\begin{eqnarray}
&&\!\!\!\!w_\alpha(x)\eql \phi_\alpha(x) +2^\alpha\mo\wu_\alpha(2x)+
2^\alpha\md\wt_\alpha(2x-1),
\label{daweql}\\
&& \!\!\!\!w_\alpha(x) \equiv \da w(x), \quad\phi_\alpha(x)\equiv \da\phi(x),
\label{defwalpha}
\end{eqnarray}
which is basically the same as \rf{weql} with $\phi$ changed into $\da \phi$
and the random multiplier rescaled by a factor $2^\alpha$.

Various theoretical results for structure functions were obtained 
for rmp's in Ref.~\cite{benzi93}. In particular it was shown that the
structure function of order $p\ge 0$ has (with our notation) the scaling exponent
\begin{equation} 
\zeta_p =-\ln_2 \la |m|^p\ra.
\label{benzizetap}
\end{equation}
 Here, we need new results for probability distributions. It turns out that
there is a class of space-independent linear random maps (Kesten maps) whose
probability distributions are very closely connected to those of the
rmp's. They are discussed in the next subsection.

\subsection{Kesten's random maps}
\label{s:kesten}

Consider the sequence of random variables defined recursively by
\begin{eqnarray}
&& w_0=1, \label{defkz}\\
&& w_{n+1}=1 +m_nw_n, \label{defkn}\\
&&n=0,1,\ldots,\nonumber
\end{eqnarray}
where the $m_n$'s are independent copies of the random multiplier $m$,
with the same definition as used above. The limit of $w_n$ for $n\to\infty$, 
if it exists, has the same distribution as
\begin{equation}
w=1+m_1+m_1m_2+\ldots +m_1m_2\cdots m_n+\ldots
\label{kestenseries}
\end{equation}
Kesten  has obtained a number of important results for a more
general class of maps where the multipliers may be random matrices 
\cite{kesten73}. In its scalar form \rf{defkn} has attracted the attention of
statistical physicists interested in disordered systems or finance 
\cite{dh83,clnp85,sornette98}.
Results on the existence of invariant measures for Kesten maps and more
general nonlinear random maps are found for example in
Refs.~\cite{jsb98,df99,blank01}. Here, we are only interested in the scalar linear map
defined by \rf{defkn}, which we shall call the ``Kesten map'' associated to
the rmp. 

We now summarize a few important results for Kesten maps without giving
proofs, but we shall give occasional hints. Let
\begin{equation}
\lambda \equiv \la \ln |m|\ra
\label{deflyapunov}
\end{equation}
be the Lyapunov exponent of the map. If $\lambda >0$ the $w_n$'s run away
to infinity almost surely. ($\ln |m_1m_2\ldots m_n|$ grows approximately
as $\lambda n$.) If $|m|\le a<1$ almost surely, then the series
\rf{kestenseries} converges to a value $<1/(1-a)$. Hence, the tail
behavior of $w$ at large values is trivial. When $\lambda <0$ and there is a
non-trivial tail, the nature of the invariant measure (the distribution of
$w$)  depends in particular
on the average slope $\la |m|\ra$ of the map. If $\la |m|\ra>1$ it may
be shown that the invariant measure is absolutely continuous, i.e. that
there is a finite probability density for $w$. (This may still hold, but only 
in special instances, for $\la |m|\ra<1$.)  The most important result is that,
for $\la |m|\ra>1$ and $\lambda <0$, there is a power-law tail:
\begin{equation}
\prob \lb|w|>\xi\rb \propto \xi ^{-\ps}, \quad \xi \to  \infty,
\label{powerlaw1}
\end{equation}
where $\ps$ is the single positive number such that
\begin{equation}
\la |m|^{\ps}\ra =1.
\label{moment1}
\end{equation}
(Take the absolute value of \rf{defkn}, raise it to the $p$th power, average
and neglect the contribution of the additive 1 when $p$ is close to $\ps$ and
just below, so that this moment comes predominantly from the tail behavior of 
$w$.)

Now we give an alternative derivation of this power-law behavior, based on
large deviations, which is in spirit very close to the phenomenological
derivation of power laws in \xsc{s:pheno}. For background on large deviations
see, e.g., Refs.~\cite{varadhan84,ellis85} (an elementary introduction may be
found in Section~8.6.4 of Ref.~\cite{frisch95}).

Consider the $n$th term on the r.h.s. of \rf{kestenseries}. Since the $m$
variables have symmetric distributions, it may written
\begin{equation}
\pm |m_1||m_2|\cdots |m_n| = \pm 2^{-(a_1+a_2+\ldots +a_n)}, 
\quad a_i\equiv -\ln_2|m_i|,
\label{defai}
\end{equation}
where the plus and minus signs are taken with equal probabilities 1/2. 

First, let us assume that just one term dominates the tail behavior
of $w$:
\begin{equation}
w \approx \pm |m_1||m_2|\cdots |m_n|
\label{cheat}
\end{equation}
 (or, more precisely, that  the right scaling is obtained; we shall
come back to this later). Large deviations theory (expressed here somewhat
loosely) tells us that, for large $n$ and given $h<0$,
\begin{equation}
\prob \lb a_1+a_2+\ldots +a_n \approx nh \rb \sim 2^{-nF(h)},
\label{looselarge}
\end{equation}
where $F(h)$ is the Cram\'er (or rate) function, defined here to be positive
to ensure consistency with the definition of the codimension used
in \xsc{s:pheno} (this is also why we use powers of 2 and the minus sign in the
definition of $a_i$).

From \rf{cheat} and \rf{looselarge}, and after integration over all
possible $h$'s, we obtain
\begin{eqnarray}
&&\prob \lb |w|> \xi\rb \propto \xi ^{-\ps}, \label{wgxi}\\[1ex]
&&\ps = \inf_{h<0}{F(h)\over -h}.
\label{againforalphz}
\end{eqnarray}
Similarly, for $p\ge 0$, we find
\begin{eqnarray}
&&\la \left(|m_1||m_2|\cdots |m_n|\right)^p\ra=2^{-n\zeta_p},
\label{kestensf}\\[1ex]
&&\zeta_p =-\ln_2\la |m|^p\ra = \inf_h \left(ph+F(h)\right).
\label{kestenlegendre}
\end{eqnarray}
Note that \rf{kestenlegendre} is just Cram\'er's result that $F(h)$ is the
Legendre transform of the logarithm of the characteristic function of the
variables $a_i$.  Note also that \rf{wgxi}, \rf{againforalphz}, \rf{kestensf}
and \rf{kestenlegendre} resemble \rf{powerlaw},
\rf{pstar}, \rf{defsp} and \rf{zetap} established in \xsc{s:pheno}.
It is easily shown by manipulations resembling those of \xsc{s:pheno} that
 $\ps$
has exactly the value predicted by \rf{moment1} and that the minimum
in \rf{againforalphz} is achieved at
\begin{equation}
h_\star = -\la |m|^\ps \ln_2|m|\ra.
\label{kestenhs}
\end{equation}

Obviously, in order for $\ps$ to exist, negative values for $h$ must be
permitted or, in other words,  the variable $|m|$ must be allowed to take 
values in excess of unity. 

In evaluating the tail of the series \rf{kestenseries} we have so far assumed
that it suffices to take a single term of suitable order $n$. Two questions
arise: (i) how large should $n$ be and (ii) what happens if we take the whole
series? The answer to the former follows from \rf{defai}, \rf{cheat} and
\rf{looselarge}: since $|w| \approx 2 ^{-nh_\star}$, we need to have $n$ of
the order of $\ln_2|w|/|h_\star|$. We observe that, the larger $|h_\star|$,
the faster (in $n$) large $|w|$-values will be reached. Consistently with the
definition given in \xsc{s:pheno} we call $C \equiv |h_\star|$ the {\it
multifractality parameter}. For the latter question we just briefly report in
words our findings which are a bit too technical for the present paper: each
of the terms in \rf{kestenseries} is, except for a random sign factor, an
exponential of a sum of $a_i$ terms. These sums perform a random walk. If, for
large $n$, we put the condition that there should be a large deviation such
that the slope of the walk is approximately $h\ne \la a\ra$, then the walk is
itself close to a straight line of slope $h$ (with small Gaussian fluctuations
around it).  One then finds the same scaling result as given by \rf{wgxi} and
it may be checked that the inclusion of more than one term just affects the
constant in front of the power law.

\subsection{Multifractal properties of the random multiplicative process}
\label{s:multirmp}

The fact that the rmp's, as defined at the beginning of \xsc{s:rmp}, are
multifractal has already been proved in Ref.~\cite{benzi93}. The $\zeta_p$
function given there (our eq.~\rf{benzizetap}) is precisely the same as
\rf{kestenlegendre} for the Kesten map. This is, of course, not accidental
since the rmp was introduced by analogy with  random multiplicative 
cascade models, for
which a simple relation is known to exist between their multifractal
properties and the large deviations of the logarithms of the multipliers (see,
e.g., Section~8.6 of Ref.~\cite{frisch95}).

Now we turn to tails of the fractional derivatives of rmp's and show that they
have precisely the properties conjectured in \xsc{s:pheno}. We shall limit
ourselves to instances where the mean slope of the associated Kesten map is
greater than one or, equivalently using \rf{moment1}, $\ps <1$.
Let us make a preliminary remark
regarding rmp's for which the largest multiplier is exactly unity, as is the
case of the example considered in Section~4 of Ref.~\cite{benzi93}, when
account is made for a factor $\sqrt 2$ in equation~(9) of that reference. In
such a case, neither the associated Kesten map nor the rmp itself can have a
power-law tail for $w$: if the multiplier 1 has finite probability, say
$\rho<1$, after $n$ iteration, $w$ is at most $O(n)$ with probability $\rho
^n$. If we now take a fractional derivative of order $\alpha>0$, we have
already noted that this rescales the multipliers by a factor $2^\alpha$ so
that power-law tails are no more ruled out. In the Kesten map associated to
$\da w$ there will be a power-law tail with exponent $-\ps$ given, in view of
\rf{moment1}, by
\begin{equation}
\la 2^{\ps\alpha}|m|^\ps\ra =1.
\label{momentalpha}
\end{equation}
By \rf{kestenlegendre} this is equivalent to $\zeta_\ps =\ps\alpha$, which is
identical to \rf{zetappalpha}. For the full rmp we do not know a rigorous
derivation of this relation but we shall now show that if $\da w(x)$ has, for
all $x$, a power-law tail probability with exponent $-\ps$ ($0<\ps<1$), then
$\ps$ is given by \rf{momentalpha} and the same exponent applies to the global
homogeneous process $\da u$.

For this, following rather closely an argument given
in Ref.~\cite{tst97}, consider
\begin{equation}
K_\alpha(z,x) \equiv \la e ^{\imath z \da w(x)}\ra,
\label{defkzx}
\end{equation}
the characteristic function of the fractional derivative of the  local rmp.
By \rf{daweql} and the independence of $\mo$, $\md$, $\wu$ and $\wt$,  the
following integral equation is obtained 
\begin{eqnarray}
K_\alpha(z,x) = \la e ^{\imath z \da \phi(x)}\ra 
&&\int_\rset K_\alpha(2^\alpha y,2x)p(y)\,dy  \nonumber\\
&&\times\int_\rset K_\alpha(2^{\alpha} y,2x-1)p(y)\,dy,
\label{intequ}
\end{eqnarray}
where $p(\cdot)$ is the probability density of $m$. By the assumption made, 
there is a small-$z$ expansion of the form:
\begin{equation}
K_\alpha(z,x) = 1+|z|^\ps g_\alpha(x) + {\rm h.o.t.},
\label{expandK}
\end{equation}
where h.o.t. (higher order terms) stands for $o(|z|^\ps)$. Expanding
\rf{intequ}, collecting all $O\left(|z|^\ps\right)$ terms and integrating
overs $x$, we find that
\begin{equation}
g_\alpha = g_\alpha \int_\rset 2 ^{\ps\alpha}|y|^\ps p(y)\,dy,
\label{welllll}
\end{equation}
where $g_\alpha \equiv \int_\rset g_\alpha(x)\,dx$. 
Eq.~\rf{momentalpha} follows immediately from \rf{welllll}. From a small-$z$
expansion of \rf{good},
with $u$ and $w$ replaced by $\da u$ and $\da w$, respectively,
it follows that $\da u$ has the same tail exponent $-\ps$.

Finally, we briefly indicate the changes in tail probabilities which occur
when the order of fractional differentiation $\alpha$ varies. For simplicity
we assume that $m$ is bounded and, without loss of generality, that the
largest value $m_{\rm max} =1$ (otherwise change $\alpha$ into $\alpha -\ln_2
m_{\rm max} $). For $\alpha=0$ we already noted that no power-law tail can be
present. The down convexity of the function $\ps\mapsto
2^{\alpha\ps}\la|m|^\ps\ra$ implies that \rf{momentalpha} has at most one
solution other than $\ps =0$.  Whether or not is has one depends on the
Lyapunov exponent of the Kesten map with multiplier $2^\alpha m$. If
$0<\alpha < -\la\ln_2 |m|\ra$ there is a power-law tail.  If $\alpha >
-\la\ln_2 |m|\ra$ there is runaway.

\subsection{Designing numerical experiments with the random multiplicative 
process}
\label{s:numericrmp}

\begin{figure}
\iffigs 
\centerline{\psfig{file=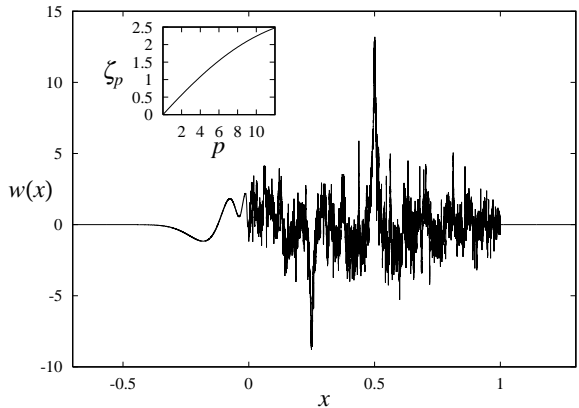,width=8cm}}
\else\drawing 65 10 {benzi-local+zetap-inset}
\fi
\vspace{2mm}
\caption{The (local) Roman rmp \protect\cite{benzi93} and its $\zeta_p$ function (as inset).}
\label{f:benzi}
\end{figure}
In this Section we perform numerical experiments with a class of rmp's, called
dyadic, where the multipliers take only two values (and their
opposites). The example considered in Section~4 of Ref.~ \cite{benzi93},
here called the {\it Roman rmp},  is an instance. It has
\begin{equation}
m =\cases{\pm 2^{-1/3}, &each with probability $7/16$;\cr
\pm 1, &each with probability $1/16$. \cr}
\label{defbenzi}
\end{equation}
The general class studied  will be 
\begin{equation}
m =\cases{\pm m_1, &each with probability $q/2$;\cr
\pm 1, &each with probability $(1-q)/2$, \cr}
\label{defgeneral}
\end{equation}
where $0<m_1<1$ and $0<q<1$.
It is easily shown that for the Kesten map associated to $u$
the  $\zeta_p$'s and the multifractality parameter are given by 
\begin{eqnarray}
&&\zeta_p = -\ln_2\left( q m_1^p +1-q\right), \label{dyadiczetap}\\
&&C(\alpha) = -{\ln_2\left( q m_1^\ps +1-q\right)\over \ps} +{qm_1^\ps\ln_2 m_1
\over qm_1^\ps+1-q},
\label{calphadyadic}
\end{eqnarray}
where $\alpha$ and $\ps$ are related by \rf{momentalpha}.
From \rf{dyadiczetap} we can calculate the minimum and maximum H\"older
exponents:
\begin{equation}
\hmin =0, \qquad \hmax = -q\ln_2 m_1.
\label{hminmax}
\end{equation}

\begin{figure}
\iffigs 
\centerline{\psfig{file=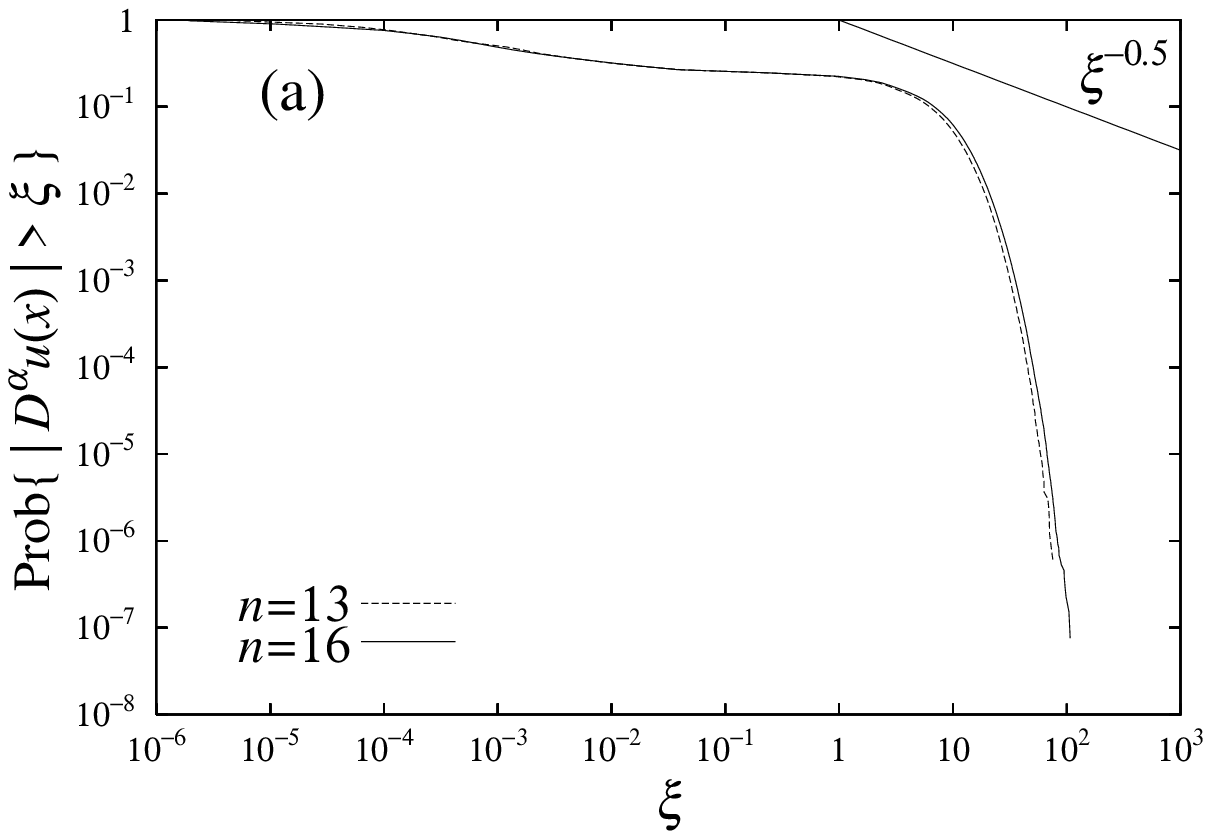,width=9cm}}
\centerline{\psfig{file=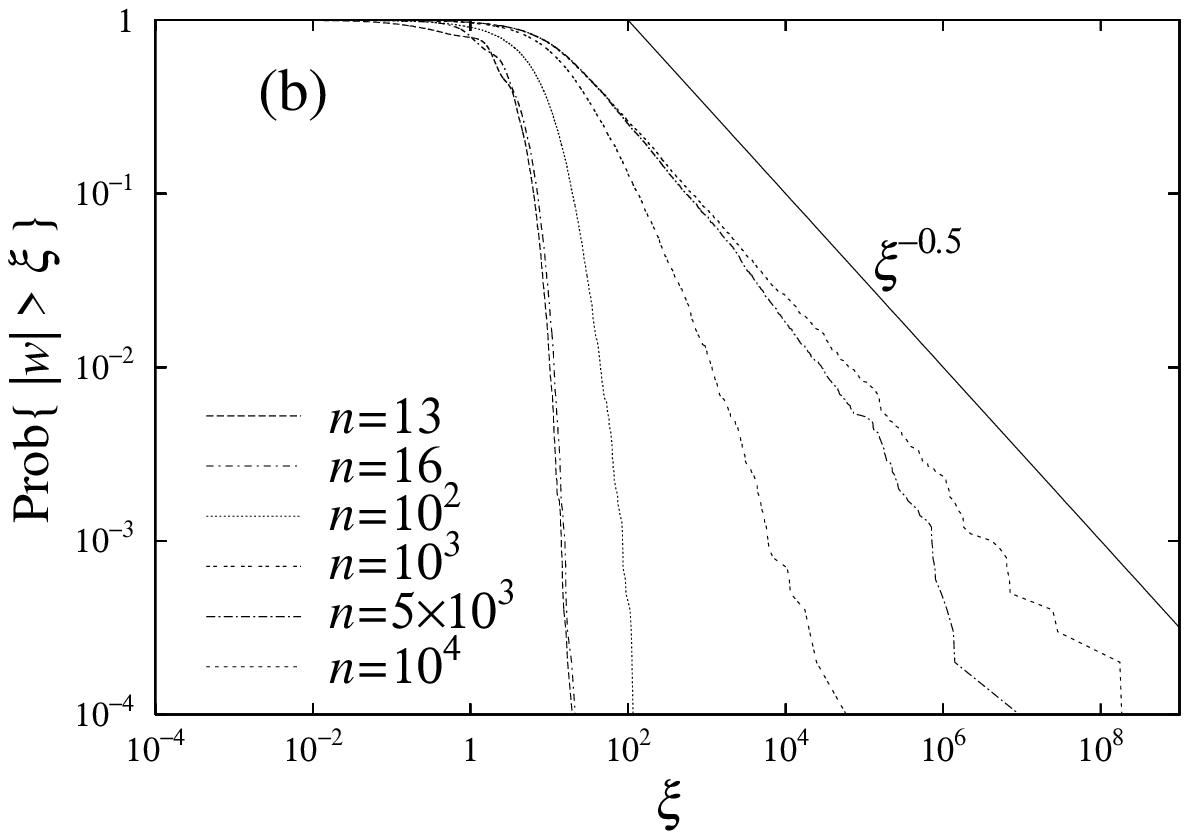,width=9cm}}
\else\drawing 65 10 {MC for Benzi and assoc. Kesten}
\fi
\vspace{2mm}
\caption{(a): distribution of fractional derivative of order $\alpha=0.2895$
($p_\star =0.5$) for the Roman rmp \protect\cite{benzi93}. Number of iterations
(levels) $n$ in the Monte-Carlo simulation with $10^2$ realizations as
labeled. The slope of the line is the predicted asymptotic behavior. No
power-law is observed. (b): distribution for the associated Kesten map
\rf{defkn} with $10^4$ realizations. Power-law behavior is observed
only when the number of iterations $n\ge 5\times 10^3$.}
\label{f:benzitail}
\end{figure}
The simulations reported hereafter have $n\le 16$ iterations of the rmp (but
much more for the Kesten maps).  The
width of the Mexican hat is always $\sigma = 0.1$. For the construction of the
local rmp, the simulation interval, from $-0.7$ to $1.3$, is chosen such that
the local rmp $w(x)$ is down to (single-precision) roundoff level at the edges
of this interval. The resolution is $2^{-(n+4)}$ (the $+4$ is needed because
the width of the Mexican hat is substantially smaller than unity). The global
rmp is constructed by adding typically one hundred independent Poisson-shifted
local realizations with a density of the Poisson distribution
$c=0.5$. Fractional derivatives are calculated as multiplications by
$|k|^\alpha$ of the (discrete) Fourier transform (windowing not needed because
of the rapid falloff at the edges).  Cumulative probabilities are calculated
from the global rmp using rank ordering to avoid binning
\cite{zipf49,skkv96}. Cumulative probabilities for Kesten maps are obtained by
Monte-Carlo simulation of the random map with a number of iterations ranging
from $n=13$ to $n=10^4$ and typically $10^4$ realizations.
\begin{figure}
\iffigs 
\centerline{\psfig{file=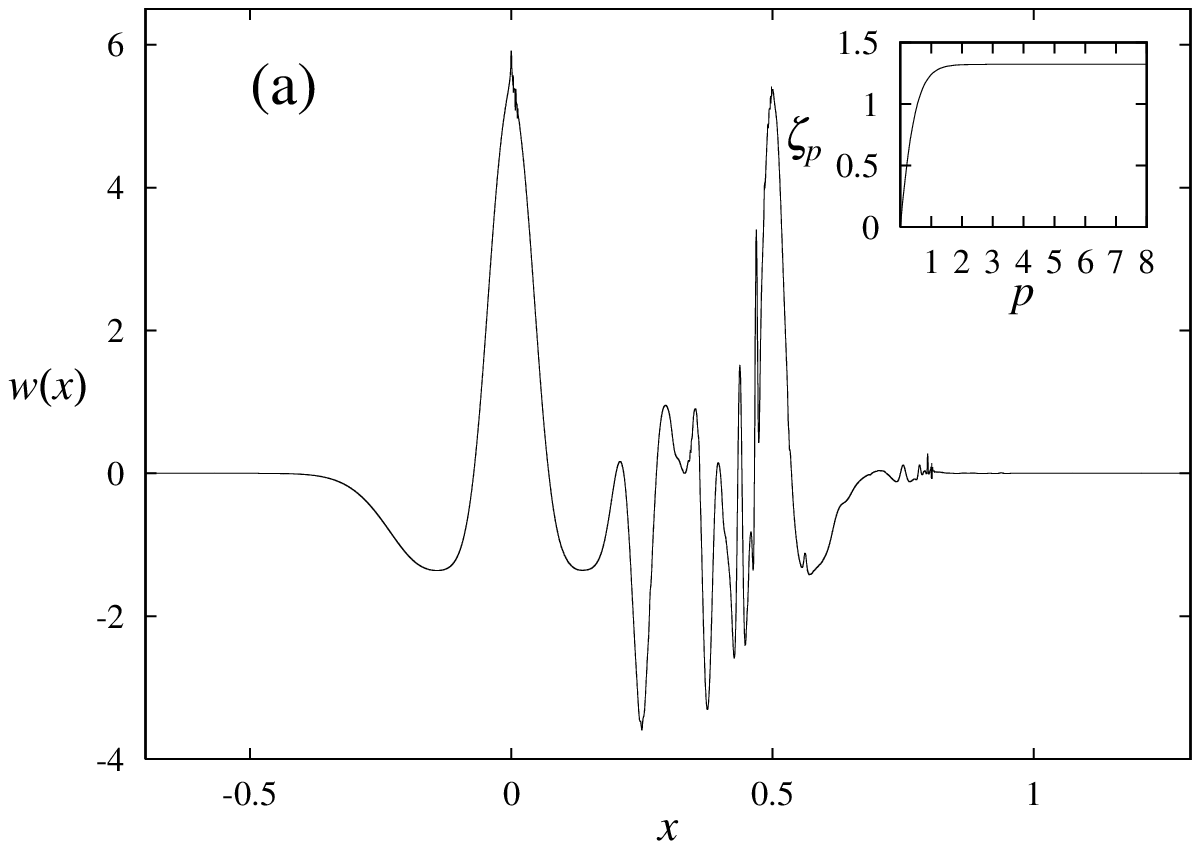,width=9cm}}
\centerline{\psfig{file=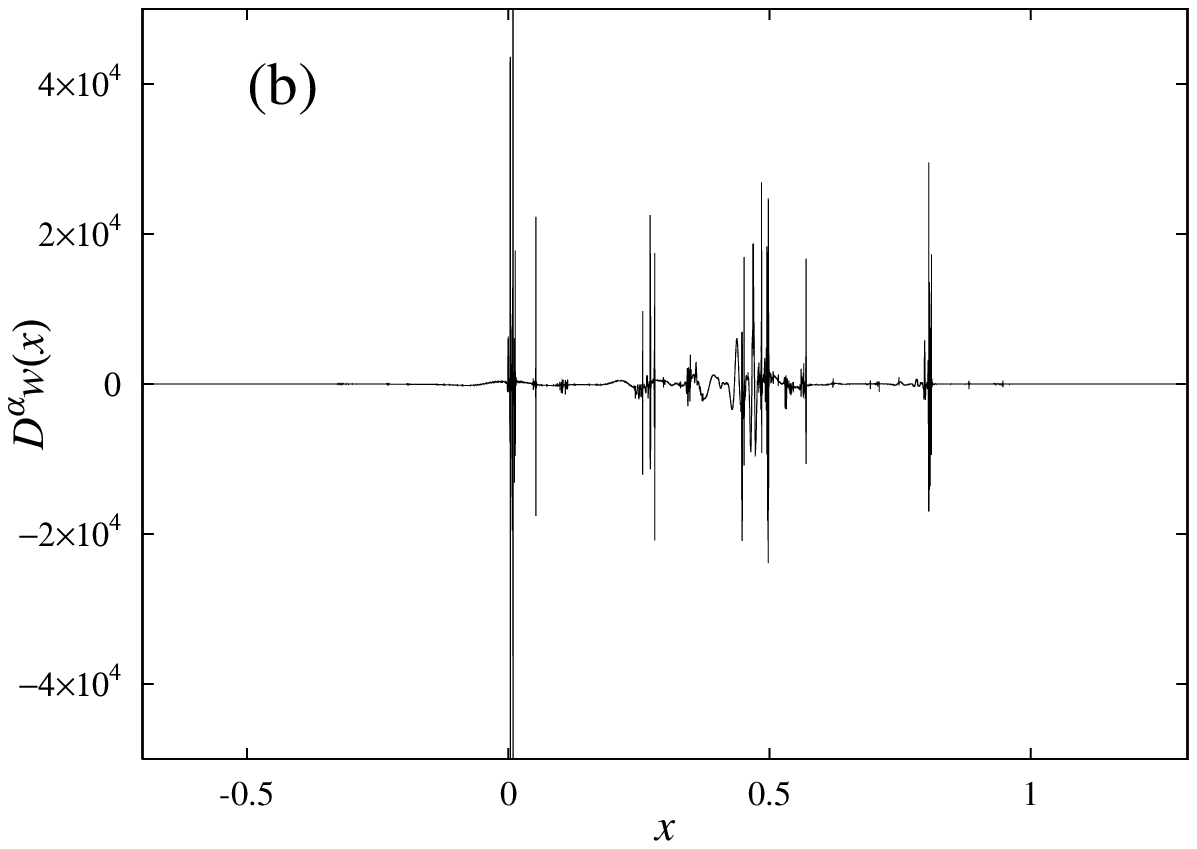,width=9cm}}
\else\drawing 65 10 {optimal-local+zetap-insert}
\fi
\vspace{3mm}
\caption{(a): the (local) optimal rmp defined by \rf{defoptimal} and its
$\zeta_p$ function (as inset).; (b): fractional derivative of order
$\alpha=1.368$ applied to this process; note the spikes.}
\label{f:optimal}
\end{figure}
\xfg{f:benzi} shows a graph of one realization of the (local) Roman rmp for
$n=13$.  This has  $\hmax = 7/24 \approx 0.29$. \xfg{f:benzitail}(a) shows the
cumulative probability of the fractional derivative for an $\alpha$ chosen
to give $\ps =1/2$ and $n=16$. No power-law tail is observed. The associated
Kesten map does reveal the power-law tail but only after about $5000$
iterations
(in the rmp this would correspond to the smallest scale being
$2^{-5000}$!).
\noindent
The reason for this slow convergence is the exceedingly  small value of the
multifractality parameter $C=0.0022991$. Hence, the Roman rmp is not a good
candidate for seeing power-law tails in fractional derivatives.
\begin{figure}
\iffigs 
\centerline{\psfig{file=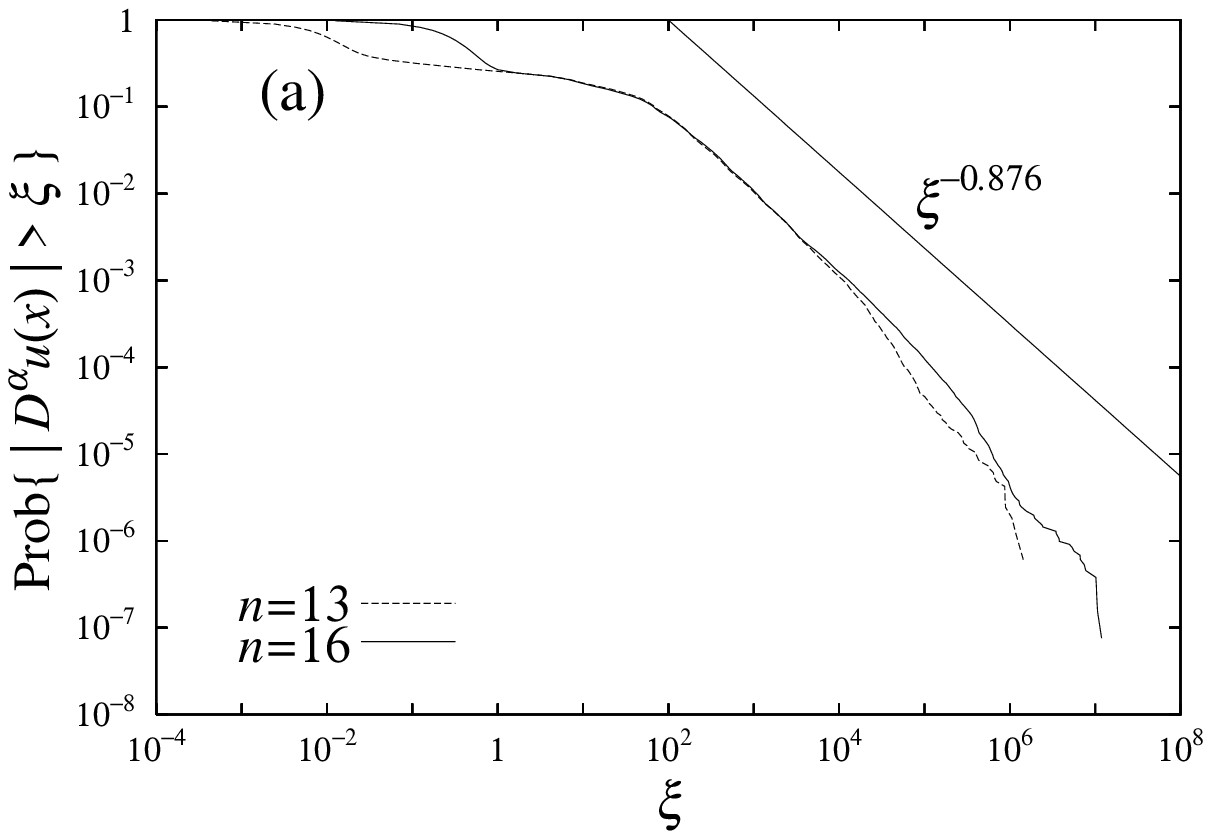,width=8cm}}
\centerline{\psfig{file=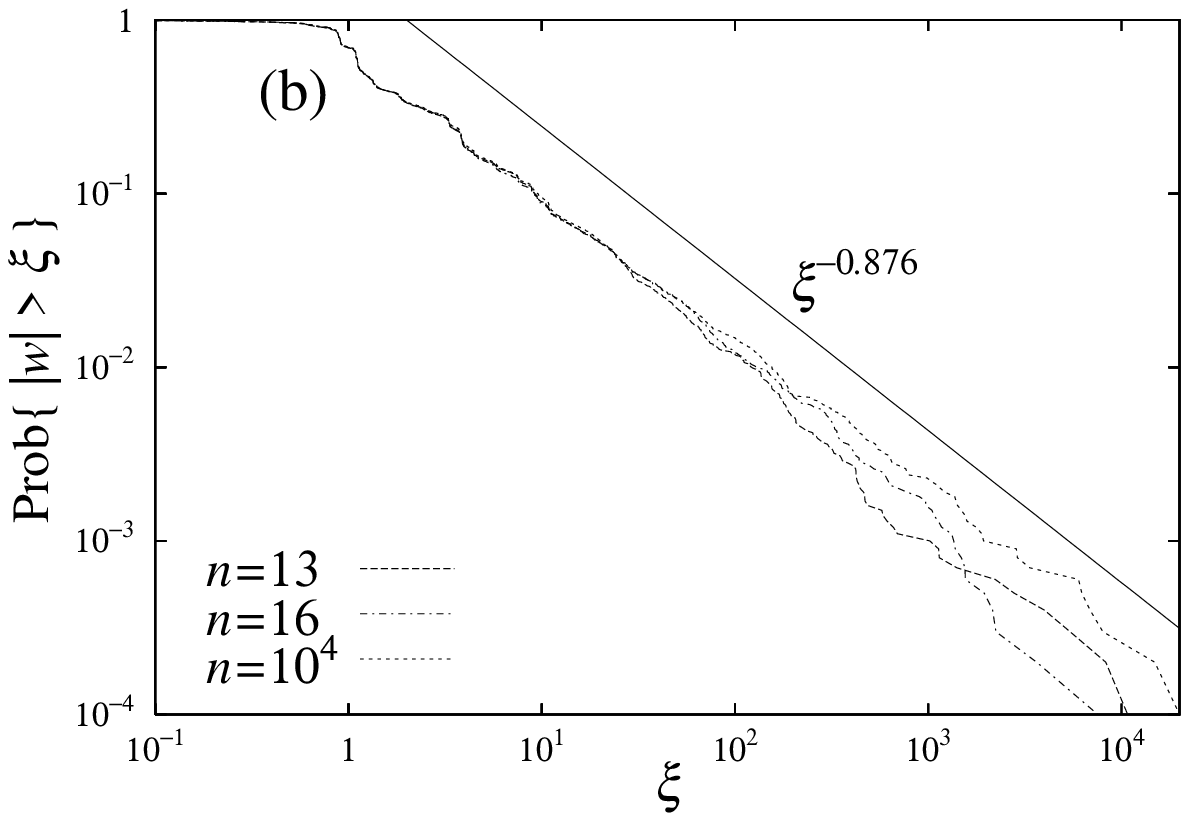,width=8cm}}
\else\drawing 65 10 {MC for optimal and assoc. Kesten}
\fi
\vspace{2mm}
\caption{As in Fig.~\ref{f:benzitail}, but for the optimal rmp
\rf{defoptimal} and $\alpha = 1.368$ ($p_\star=0.876$). Note the
conspicuous power-law tail after about 13 to 16 iterations.}
\label{f:optimaltail}
\end{figure}

For obtaining better candidates we systematically search through
parameter space, varying $m_1$, $q$ and $\alpha$ and trying to maximize the
multifractality parameter $C(\alpha)$ given by \rf{calphadyadic}. 
In practice, after
delineating the acceptable ranges of these parameters, we randomly choose
$3\times 10^4$ triplets $(m_1,q,\alpha)$ and each time calculate $C(\alpha)$.
We thus obtain about half a dozen triplets for which $C(\alpha)$ is close
to unity, i.e. more than 400 times larger than the value for the Roman rmp.
We then select the one for which the Kesten map displays a conspicuous
power-law tail for $n$ as low as possible. This turns out to be $n\approx 13$
(around $n\approx 10$ a power-law tail begins to emerge).
We call this the ``optimal rmp''. Its parameters are:
\begin{equation}
m_1=0.04, \quad q=0.6, \quad \alpha=1.368, \quad C(\alpha) =0.9867.
\label{defoptimal}
\end{equation}
\xfg{f:optimal} shows the local optimal rmp and its fractional derivative of
order $\alpha=1.368$, which gives $\ps=0.876$. Note that the graph of
$\zeta_p$ (inset on \xfg{f:optimal}(a)) is much more curved than that for the
Roman rmp.  \xfg{f:optimaltail}(a) shows the cumulative probability which is
seen to display a power-law tail already for $n=13$ and $n=16$. Note that the
cumulative probability for the associated Kesten map \xfg{f:optimaltail}(b) is
a somewhat wiggly power law due to the presence of 
log-periodic subdominant corrections \cite{jsb98}; this
phenomenon is less pronounced if we use lower values of $p_\star$
(e.g. 0.5, not shown).

\section{Analysis of experimental data}
\label{s:data}
\begin{figure}
\iffigs 
\centerline{\psfig{file=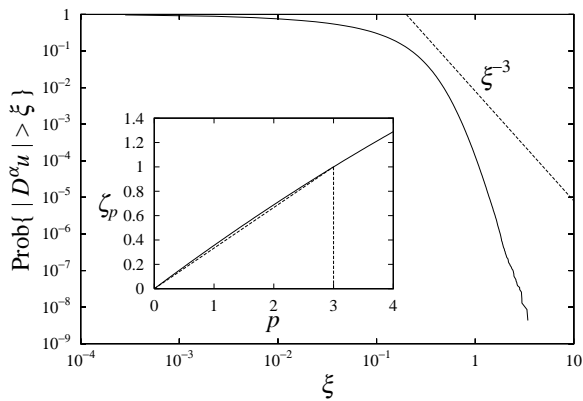,width=8cm}}
\else\drawing 65 10 {P(w) for Modane $p_\star=3$. $\zeta_p$ inset}
\fi
\vspace{2mm}
\caption{Distribution of fractional derivative of order $\alpha=1/3$ for the
Modane turbulence data. The power law with exponent $-p_\star=-3$ is the
predicted tail behavior at infinite Reynolds number. In fact, no scaling
region is observed. Inset: corresponding $\zeta_p$ graph which is close to its
Kolmogorov 1941 value, a straight line of slope $1/3$.}
\label{f:modanep3}
\end{figure}

As already stated in the Introduction, multifractality was introduced
to interpret experimental turbulence data, specifically the up convex
bending of the graph of $\zeta_{p}$.  Some of the best high-Reynolds
turbulence data have been collected at the ONERA wind tunnel by the
Grenoble group from LEGI. The Taylor-scale based Reynolds number is
$R_{\lambda} \approx 2700$ with about three decades of scaling (for
definitions see, e.g., Ref.~\cite{frisch95}).  We  use
(longitudinal) velocity data obtained from this group to check for the
possible presence of power-law tails in fractional derivatives.  A
total of $2.3\times 10^{9}$ data points (in 12 segments of variable
lengths) is analyzed.  To calculate the fractional derivatives of these
non-periodic data we use Hann windowing (see, e.g., Section 13.4 of
Ref.~\cite{press92}).  The different segments are processed
separately and the union of all fractional derivative data is then
rank ordered to produce cumulative probabilities (using just a single
segment does not produce significantly different results; as we shall
see the problem we shall encounter has little to do with noise which
can be reduced by using more data).  

\xfg{f:modanep3} shows the
probability of the fractional derivative of order 1/3, a value chosen
because, by Kolmogorov's four-fifths law, the corresponding $\ps$
should be exactly three \cite{frisch95}.  \xfg{f:modanephalf} shows
the case $\alpha = 13/36$, which corresponds to $\ps =1/2$ (when 
using the lognormal law $\zeta_{p}=(p/3)+(\mu/18)p(3-p)$ with $\mu=0.2$,
an excellent approximation for $p$ up to at
least six). For both instances we show the nearly straight graph of 
$\zeta_{p}$ over the relevant range of $p$'s. 
In neither case do we see any trace of the predicted 
power law tail.  

\begin{figure}
\iffigs 
\centerline{\psfig{file=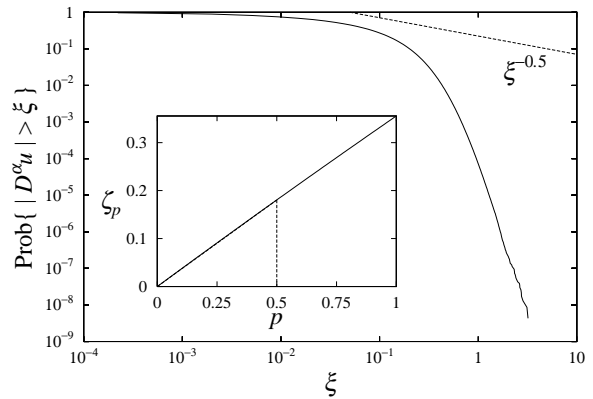,width=8cm}}
\else\drawing 65 10 {P(w) for Modane $p_\star=1/2$. $\zeta_p$ inset}
\fi
\vspace{2mm}
\caption{As in Fig.~\ref{f:modanep3} but for $\alpha =13/36$ 
($p_\star \approx 1/2$).}
\label{f:modanephalf}
\end{figure}

Here, a digression is in order. The fractional 
derivative analysis of this turbulence data
was actually made by us before the work on random
multiplicative processes reported in \xsc{s:rmp}. The latter was 
motivated in part by our desire to interpret the failure to see
power laws in experimental data.  Of course, now it is clear why
we do not see such power laws. The multifractality parameter of the
experimental data, using the lognormal approximation, is
\begin{equation}
 C(\alpha)=  {2+\mu\over 6} -\alpha.
\label{Clognormal}
\end{equation}
For $\alpha = 1/3$ we have $C\approx 0.033$ while for $\alpha=13/36$ 
we have $C\approx 0.0055$.
These values are far too small for power-law tails 
to be observable with a three-decade 
long inertial range. Actually, what we observe in \xfg{f:modanep3} and 
\ref{f:modanephalf} at high values of the modulus $\xi$  of the 
fractional derivative is due to the viscous cutoff and can be 
predicted by the argument of Ref.~\cite{fs91}, adapted
to fractional derivatives. 

Higher Reynolds number data with four decades of scaling will soon
become available thanks to the use of large-scale cryogenic Helium
facilities \cite{gagne01}.  This increase is unlikely to be sufficient
to see the predicted power laws.  Alternative methods, based on the
processing of the local dissipation measure by fractional derivatives
or integrals should also be explored.

We now consider financial data for which the observation of power-law tails
for fractional derivatives could be easier, although the data are far scarcer
than in turbulence.  Stock market indices are generally believed to display
multifractal scaling.  We refer, for example, to
\hbox{http://www.ncrg.aston.ac.uk/$\sim$vicenter/financial.html} which
contains a number of references to multifractal (or spuriously multifractal)
behavior of financial time series.  We use S~\&~P Futures intraday
(minute-by-minute) data with linear interpolation for the approximately 12 \%
of missing data points due to lack of activity.  The total number of points
including interpolated data is $1.88\times 10^6$.  We analyze their logarithms
(because returns are defined as differences of logarithms).  Hann windowing is
used again (higher-order windowing makes no substantial difference.).
Structure functions (not shown) of order up to $p=4$ display almost three
decades of scaling.  The corresponding $\zeta_p$ function is shown as an inset
on \xfg{f:sandp}. For $\alpha = 0.348$ we show the cumulative probability (by
rank ordering) in \xfg{f:sandp}.  The $\zeta_p$ graph is much more curved than
that for turbulence data.  The multifractality parameter, $C(0.348)\approx
0.30$, being almost one order of magnitude larger than for the turbulence
data, it is not surprising to observe a power-law range ($\propto
|\xi|^{-3.5}$) in the cumulative probability.  Let us finally point out that,
in one respect, such financial data are closer to Burgers turbulence (which
easily displays power-law tails for fractional derivatives) than to
Navier--Stokes turbulence.  Indeed, there are occasional strong
discontinuities, such as financial crashes.
\begin{figure}
\iffigs 
\centerline{\psfig{file=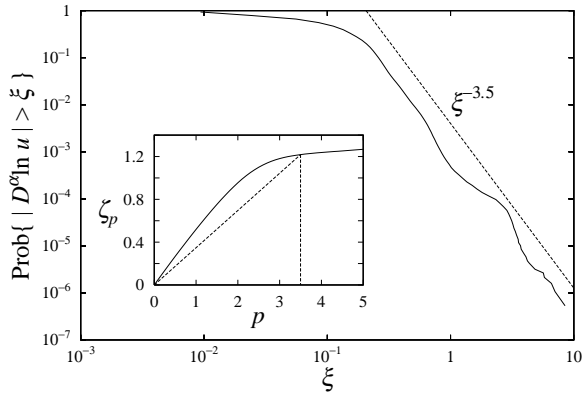,width=8cm}}
\else\drawing 65 10 {financial $P(w)$ with zetap inset}
\fi
\vspace{2mm}
\caption{Distribution of fractional derivative of order $\alpha=0.348$
($p_\star=3.5$) for S \& P 500 Futures intraday financial data
(minute-by-minute, 1982 April 21 -- 1999 February 26). The bump around $\xi = 2$ is
 mainly coming from the vicinity of the crash of October 1987.
 Inset: corresponding $\zeta_p$ graph and line of slope 
$\alpha$. }
\label{f:sandp}
\end{figure}

\section{Conclusion}
\label{s:conclusion}

We have shown in this paper that multifractality of a function has a
signature in the tail behavior of probabilities of fractional
derivatives.  This signature may frequently be invisible, since
power-law tails emerge only if the number of scaling octaves times the
multifractality parameter exceeds a value found empirically to be
around 10.  We recall that the multifractality parameter $C$ is a measure
of how strongly the functions departs from being self-similar. 
Incompressible three-dimensional turbulence having $C\approx 1/30$ is 
not a good candidate for seeing such tails. There are many other 
turbulence-like problems displaying multifractality, such as passive or
active scalars and magnetohydrodynamic turbulence. Most of these
should have larger $C$'s.

It is known that the standard way of detecting multifractality, by
structure functions, can  produce spurious multifractality.
One reason is that self-similarity does already produce structure 
functions with scaling, albeit with $\zeta_{p}$ linearly proportional 
to the order $p$. If there is a mechanism, such as contamination by
subdominant terms \cite{aflv92} or very slow convergence 
\cite{bbf94,bcpv97}, which slightly changes the apparent value of $\zeta_{p}$,
a self-similar function may look multifractal.  
Such a phenomenon is ruled out with our new method which 
detects only very robust forms of multifractality. 

We mention here some open mathematical issues.  Jaffard has recently
shown that multifractality is a (topologically) generic phenomenon in
many function spaces of the Sobolev family such as Besov spaces,
irrespective of any underlying dynamics \cite{meyer93}\cite{jaffard00}.
Can such a proof be extended to
show genericity of power-law tails?  For random multiplicative
processes it should be possible to give a rigorous proof of some of
the results presented somewhat heuristically in \xsc{s:rmp}.  We
notice also that the functional random map formulation \rf{weql} can
be used to investigate further the statistics of individual Fourier
modes, studied in Ref.~\cite{bp01}.

Finally, returning to three-dimensional turbulence we cannot resist
asking: since experimentally this turbulence does not ``test
positive'' when looking for power-laws tails in fractional derivatives
does it have other features which would reveal themselves only at
vastly larger -- and therefore perhaps inaccessible -- Reynolds
numbers?  Could it be that the velocity itself (without
taking any derivative) or that velocity increments have power-law tails in
their probabilities, as suggested for example in Ref.~\cite{sl84}?  This
cannot be ruled out completely.  For example, if we accept the
lognormal model (with $\mu \approx 0.2$) also for large values of $p$ (for
which the exponent $h$ is negative),
we find that $\zeta_{p}$ vanishes for $p\approx 33$.  This implies a
power-law tail with exponent $-33$.  The multifractality parameter is
$C\approx 0.37$.  Such an algebraic tail, if it exists, would require
about 30 octaves or 9 decades of scaling, that is $R_{\lambda}$'s of
about eighteen millions.  Getting there requires the crossing of a
substantial ``turbulence desert'', but a small one in comparison with
the great high-energy desert predicted by Giorgi's and 
Glashaw's grand unified theory \cite{gg74}.

\vspace*{5mm}
\par\noindent {\bf Acknowledgements}
\vspace{3mm}

We are grateful to E.~Aurell, L.~Biferale, M.~Blank, K.~Khanin, G.~Molchan,
G.~Parisi, K.~Schneider, D.~Sornette, B.~Villone and an anonymous 
referee for very useful
discussions and remarks.  Special thanks are due to J.~Bec for a remark concerning
singularities and power-law tail distributions.  Experimental turbulence data
obtained at ONERA Modane were kindly provided by Y.~Gagne and Y.~Mal\'ecot.
Part of this work was done while the authors were visiting the Department of
Applied Mathematics of the University of Porto.  Computational resources were
provided by the Yukawa Institute (Kyoto).  This work was also supported by the
European Union under contract HPRN-CT-2000-00162, by the Indo-French Centre
for the Promotion of Advanced Research (IFCPAR~2404-2) and by the Japan
Scholarship Foundation.

\end{document}